\documentclass[conference]{IEEEtran}
\IEEEoverridecommandlockouts

\usepackage{cite}
\usepackage{amsmath,amssymb,amsthm,amsfonts,multicol, nccmath}
\usepackage{multirow, ragged2e, balance}
\usepackage{graphicx, booktabs, xcolor}
\usepackage[hidelinks]{hyperref}
\usepackage{subcaption}
\usepackage{booktabs}
\usepackage{float, array, tabularray, diagbox}

\begin{document}

\title{Towards Deep Learning-Based Linear Assignment Problem Solvers for Multi-Object Tracking}

\author{\IEEEauthorblockN{Chu Tam Vu\IEEEauthorrefmark{2}, Tran Thi Thuy Quynh\IEEEauthorrefmark{2}, and Do Hai Son\IEEEauthorrefmark{3}\IEEEauthorrefmark{1}.
    \vspace{0.2cm}\\}
    \IEEEauthorrefmark{2} VNU University of Engineering and Technology, Hanoi, Vietnam. \\
    \IEEEauthorrefmark{3} School of Electrical Engineering, Computing and Mathematical Sciences, Curtin University, Australia.
    \thanks{Corresponding author: Do Hai Son (haison.do@postgrad.curtin.edu.au).}
}

\maketitle

\begin{abstract}
This paper proposes and evaluates deep learning (DL)-based linear assignment problem (LAP) solvers as a replacement for the classical Hungarian algorithm in multi-object tracking (MOT) systems. Three representative neural models are studied, namely a convolutional network (CNN), a graph neural network (GNN), and the graph-based linear assignment network~(GLAN). A cost-matrix embedding is introduced so that a single model operates on cost matrices of any size, as required in MOT where the number of objects varies over time, without training a separate model for each input shape. Extensive simulations show that the embedding improves the accuracy of the neural solvers (i.e., CNN and GNN), while GLAN attains the best accuracy among them. These solvers are then integrated into a generalized labeled multi-Bernoulli~(GLMB) tracker and evaluated on the WILDTRACK dataset, where GLAN yields tracking accuracy close to the Hungarian reference and comparable to recent DL-based trackers.
\end{abstract}


\section{Introduction}\label{sec:Intro}

Multi-object tracking (MOT) estimates the trajectories of a time-varying number of objects, and it underpins applications such as intelligent surveillance, autonomous driving, and mobile robotics~\cite{Wenhan2021, Guochen2025}. At each time step, an MOT system detects the objects and maintains their identities. The latter relies on data association, which assigns the measurements from a sensor to the objects being tracked~\cite{rakai2022data}. Tracking frameworks such as the generalized labeled multi-Bernoulli~(GLMB) filters perform this association through multiple hypotheses to estimate object trajectories over time~\cite{vo2024overview}. Each such association is commonly formulated as a linear assignment problem (LAP) that seeks a one-to-one~(1-1) mapping between the two sets at minimum total cost.

Classical LAP solvers obtain the exact solution through combinatorial optimization, including the Hungarian algorithm~\cite{kuhn1955hungarian,munkres1957algorithms}, the auction algorithm~\cite{bertsekas1988auction}, and shortest augmenting-path methods~\cite{jonker1987shortest}. The Hungarian and shortest augmenting-path methods return a globally optimal mapping with a computational complexity of $\mathcal{O}(N^3)$, and their optimized implementations run fast at moderate problem sizes. This cubic cost nevertheless grows quickly with the numbers of objects and measurements and becomes a bottleneck in large-scale or high-density tracking scenarios. Faster suboptimal alternatives, such as greedy assignment~\cite{naiem2013optimality_speed}, lower the cost by committing to locally favorable matches at the expense of the global optimality guarantee. However, both exact and greedy solvers are also discrete and non-differentiable, which limits their use in gradient-based learning and has motivated differentiable relaxations of the assignment. The Sinkhorn procedure alternately normalizes the rows and columns of a positive matrix to approximate a permutation by a doubly stochastic matrix~\cite{sinkhorn1964relationship}, and entropy-regularized optimal transport and Gumbel-Sinkhorn extend this idea to differentiable optimization and latent permutation learning~\cite{cuturi2013sinkhorn,mena2018gumbel_sinkhorn}. These methods produce soft assignments that still require a discretization step to yield a valid mapping.

Recently, deep learning (DL) networks have been used to solve the LAP~\cite{Emami2020}, replacing the iterative combinatorial search with a single forward pass. Additionally, their continuous output is differentiable with respect to the costs. Hence, the DL-based Hungarian has been exploited to train end-to-end (E2E) DL-based trackers~\cite{xu2020dhn}. Lee et al.~\cite{lee2018deep_lsap} first cast the assignment of a measurement to each object as a classification problem solved by a convolutional network (CNN) that takes the cost matrix as a two-dimensional input. In 2024, Aironi et al.~\cite{aironi2024graph} represent the LAP as a bipartite graph and apply a graph neural network (GNN), and Liu et al.~\cite{liu2022glan} propose the graph-based linear assignment network (GLAN), which learns assignment decisions from edge representations. An MOT system requires a solver that operates on cost matrices of varying size, since the numbers of objects and measurements change over time. However, the CNN~\cite{lee2018deep_lsap} and GNN~\cite{aironi2024graph} assume a pre-defined cost-matrix shape and cannot accept such variable-size inputs. GLAN, in contrast, operates on the graph of the current matrix and already accommodates any size.

To address the limitation of CNN and GNN methods, this paper introduces a cost-matrix embedding that maps any cost matrix whose dimensions do not exceed a target size $N$ into a common $N\times N$ representation. Thus, networks with a pre-defined input shape (i.e., CNN and GNN) are trained once and then applied to the variable-size problems that arise in tracking. The embedding pads a rectangular cost matrix to a common dimension before inference and projects the prediction back to the original size, leaving the network architecture unchanged. The three DL-based solvers, with the embedding applied to CNN and GNN, are studied under variable-size LAP settings and compared against the Hungarian algorithm on a large synthetic dataset in terms of assignment accuracy and inference time. All three solvers are then integrated into a GLMB tracker~\cite{Vo2017} on the real-world WILDTRACK dataset, where they replace the Hungarian assignment solver. The main contributions are as follows:
\begin{itemize}
    \item A cost-matrix embedding that enables neural LAP solvers restricted to a single input size to process cost matrices of varying size.
    \item Extensive simulations on synthetic LAP instances, together with an evaluation within a GLMB tracker on the WILDTRACK dataset.
\end{itemize}

The remainder of this paper is organized as follows. Section~\ref{sec:Background} reviews the LAP and the Hungarian algorithm. Section~\ref{sec:Proposed} presents the embedding scheme and the DL models. Section~\ref{sec:Results} reports the synthetic and tracking results, and Section~\ref{sec:Conclusion} concludes the paper.

\section{Background}\label{sec:Background}

This section reviews the LAP formulation and the Hungarian algorithm used to generate the ground-truth assignments.

\subsection{Linear assignment problem}

In MOT systems, data association is an LAP between the set of objects and the set of measurements. Each candidate measurement-to-object pair carries a cost, and a solution is a 1-1 mapping $\theta$ in which every object and every measurement appear in at most one pair. With $m$ objects and $n$ measurements, such a mapping is feasible if it contains exactly $\min(m,n)$~pairs, and we denote by $\Theta$ the set of all feasible mappings.

For the square case $m=n=N$, consider a cost matrix $\mathbf{C}\in\mathbb{R}^{N\times N}$, where $c_{ij}$ denotes the cost of assigning the $i$-th object (in the existing set of objects) to the $j$-th measurement (in the set of measurements). A mapping $\theta$ is represented by a binary assignment matrix $\mathbf{X}\in\{0,1\}^{N\times N}$, where
\begin{equation}
    x_{ij}=
    \begin{cases}
        1, & \text{if the pair }(i,j)\text{ is selected},\\
        0, & \text{otherwise}.
    \end{cases}
    \label{eq:assignment_variable}
\end{equation}

The LAP is formulated as
\begin{equation}
    \min_{\mathbf{X}}\ \sum_{i=1}^{N}\sum_{j=1}^{N} c_{ij}x_{ij},
\label{eq:lsap}
\end{equation}
subject to, for all $i,j=1,\ldots,N$,
\begin{equation*}
    \sum_{j=1}^{N}x_{ij}=1,\qquad \sum_{i=1}^{N}x_{ij}=1,\qquad x_{ij}\in\{0,1\}.
\end{equation*}
The objective in Eq.~\eqref{eq:lsap} minimizes the total cost of the selected pairs, and the constraints force each element of one set to match exactly one element of the other. Every feasible solution is therefore a permutation matrix, which represents a feasible mapping~$\theta\in\Theta$.

\subsection{Hungarian algorithm}
\begin{figure*}[t]
    \centering
    \includegraphics[width=\linewidth]{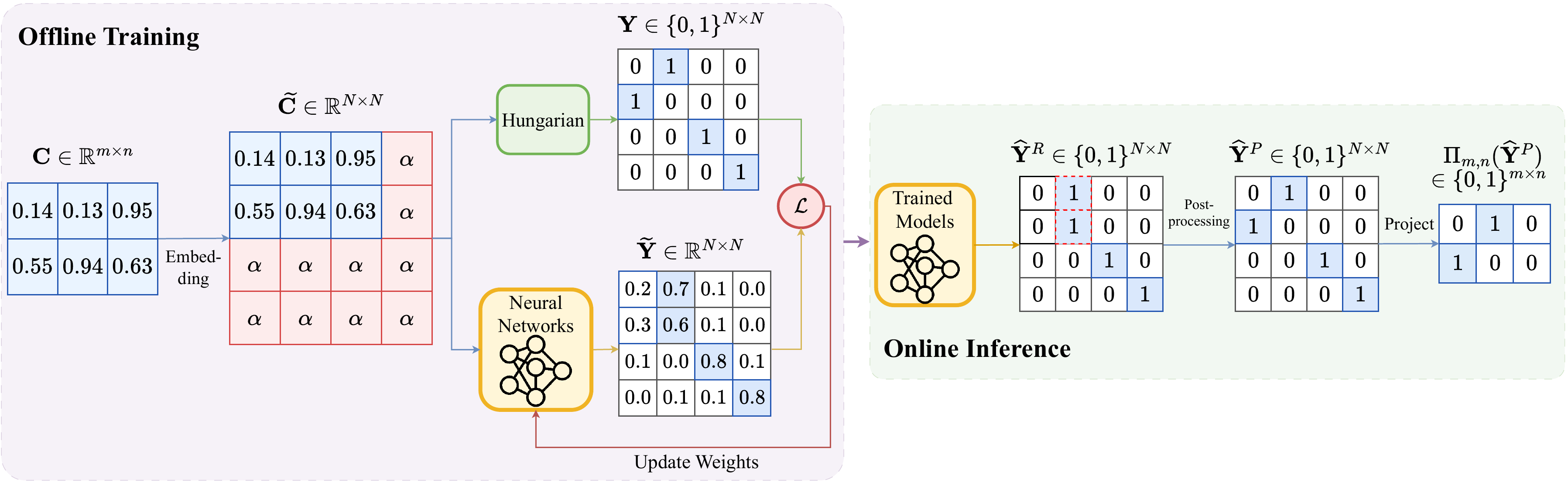}
    \caption{Cost-matrix embedding, training, and inference pipeline of the DL-based LAP solvers, illustrated with the discretization and post-processing rules of the CNN~\cite{lee2018deep_lsap}.}
    \label{fig:pipeline}
\end{figure*}
The Hungarian algorithm is one of the most widely used exact solvers for the LAP~\cite{kuhn1955hungarian,munkres1957algorithms}. Rather than searching over all permutations, it exploits the primal-dual structure of the assignment problem. Let $u_i$ and $v_j$ denote the dual variables associated with the row and column constraints, respectively. The dual problem corresponding to Eq.~\eqref{eq:lsap} is
\begin{equation}
    \max_{\mathbf{u},\mathbf{v}}
    \left(
    \sum_{i=1}^{N}u_i+
    \sum_{j=1}^{N}v_j
    \right),
    \label{eq:lsap_dual_objective}
\end{equation}
subject to
\begin{equation}
    u_i+v_j\leq c_{ij},
    \qquad i,j=1,\ldots,N.
    \label{eq:lsap_dual_constraint}
\end{equation}

For a dual-feasible pair $(\mathbf{u},\mathbf{v})$, the reduced cost associated with pair $(i,j)$ is
\begin{equation}
    \bar{c}_{ij}=c_{ij}-u_i-v_j,
    \label{eq:reduced_cost}
\end{equation}
where $\bar{c}_{ij}\geq 0$ follows from Eq.~\eqref{eq:lsap_dual_constraint}. Candidate assignments are selected from the zero reduced-cost entries. Iteratively updating the row and column potentials and constructing augmenting paths over these entries yields a complete assignment that satisfies the optimality conditions of the primal and dual problems. The resulting assignment is globally optimal, at a computational complexity of $\mathcal{O}(N^3)$.

In this work, the Hungarian algorithm generates the ground-truth $\mathbf{Y}$ for each cost matrix $\mathbf{C}$, and the neural networks are trained to approximate the mapping from $\mathbf{C}$ to~$\mathbf{Y}$.

\section{Proposed Method} \label{sec:Proposed}

This section presents the cost-matrix embedding and the CNN, GNN, and GLAN architectures for variable-size LAPs, summarized in Fig.~\ref{fig:pipeline}.

\subsection{Cost-matrix embedding}

Given an original cost matrix $\mathbf{C}\in\mathbb{R}^{m\times n}$, a padded matrix $\widetilde{\mathbf{C}}\in\mathbb{R}^{N\times N}$ is constructed, where the target size $N$ satisfies $N\geq\max(m,n)$. The original cost values are preserved, and the remaining entries take a padding constant $\alpha$:
\begin{equation}
    \widetilde{c}_{ij}=
    \begin{cases}
        c_{ij}, & i\leq m,\; j\leq n,\\
        \alpha, & \text{otherwise}.
    \end{cases}
    \label{eq:padding}
\end{equation}

In the experiments, the original cost values are normalized to $[0,1]$, and the padding value is set to $\alpha=10.0$. An assignment of $\widetilde{\mathbf{C}}$ with fewer than $\min(m,n)$ original pairs matches some original row $i$ to a padded column $j'>n$ and some original column $j$ to a padded row $i'>m$. Replacing $(i,j')$ and $(i',j)$ by $(i,j)$ and $(i',j')$ changes the cost of these two pairs from $2\alpha$ to $c_{ij}+\alpha$, which is lower since $\alpha$ is larger than every original cost. Hence, every optimal assignment of $\widetilde{\mathbf{C}}$ contains exactly $\min(m,n)$ original pairs, and its remaining pairs add the constant $(N-\min(m,n))\alpha$, so its restriction to the $m\times n$ block is also optimal for the original problem. The ground-truth is therefore generated by applying the Hungarian algorithm to the padded matrix, giving $\mathbf{Y}\in\{0,1\}^{N\times N}$, which supplies a one-hot target for every one of the $N$ rows.

Let $\widetilde{\mathbf{Y}}\in\mathbb{R}^{N\times N}$ denote the continuous score matrix returned for the padded input. It is discretized into $\widehat{\mathbf{Y}}^R \in\{0,1\}^{N\times N}$ and then turned into a feasible mapping $\theta\in\Theta$ by the post-processing of Section~\ref{sec:post}, giving $\widehat{\mathbf{Y}}^P\in\{0,1\}^{N\times N}$. This feasible assignment is projected back to the original $m\times n$ dimensions by
\begin{equation}
        \Pi_{m,n}\big(\widehat{\mathbf{Y}}^P\big)=\left[\,\widehat{y}^P_{ij}\,\right]_{1\leq i\leq m,\; 1\leq j\leq n},
    \label{eq:crop}
\end{equation}
which keeps its first $m$ rows and first $n$ columns. The ground-truth $\mathbf{Y}$ is cropped in the same manner, and the argument above guarantees that the result is optimal for $\mathbf{C}$. This guarantee does not extend to a network prediction. Cropping its post-processed assignment only removes pairs, so the result still satisfies the 1-1 constraints, but an original row may remain matched to a padded column while an original column is free, which leaves fewer than $\min(m,n)$ pairs in the cropped block. The embedding changes only the input representation and leaves the model architectures unchanged. A model with a single input dimension $N$ can therefore be trained on and applied to any cost matrix whose dimensions do not exceed~$N$, and the padding in Eq.~\eqref{eq:padding} applies unchanged for any choice of $N$ and any shape $m\times n$.

\subsection{Deep learning models}

This paper investigates CNN, GNN, and GLAN as matrix-based, node-based, and edge-based neural LAP solvers, respectively.

\textbf{CNN:} The CNN treats an $N\times N$ cost matrix as a two-dimensional input and decomposes the LAP into $N$ row-wise classification tasks~\cite{lee2018deep_lsap}. For row $i$, the network predicts an assignment-score vector
\begin{equation}
    \widetilde{\mathbf{y}}_{i}
    =
    \left[
    \widetilde{y}_{i1},
    \ldots,
    \widetilde{y}_{iN}
    \right]^\top,
    \qquad i=1,\ldots,N.
    \label{eq:cnn_output}
\end{equation}
Each classifier is trained against the corresponding row of the ground-truth $\mathbf{Y}$, which is one-hot by construction. The two-dimensional input shape is prescribed by the architecture, so the embedding is required whenever the original matrix is smaller than the target size $N$.

\textbf{GNN:} The GNN represents an LAP as a complete bipartite graph whose two node sets correspond to the rows and columns of the cost matrix~\cite{aironi2024graph}. The initial feature of each node is constructed from an entire row or column:
\begin{equation}
    \begin{aligned}
        \mathbf{x}_{i}^{(0)}
        &=
        [\widetilde{c}_{i1},\ldots,\widetilde{c}_{iN}]^\top,
        && i=1,\ldots,N,\\
        \mathbf{x}_{N+j}^{(0)}
        &=
        [\widetilde{c}_{1j},\ldots,\widetilde{c}_{Nj}]^\top,
        && j=1,\ldots,N.
    \end{aligned}
    \label{eq:gnn_node_features}
\end{equation}
The node representations are then updated through message passing:
\begin{equation}
    \mathbf{x}_{v}^{(k)}
    =
    \frac{1}{|\mathcal{N}(v)|}
    \sum_{w\in\mathcal{N}(v)}
    \operatorname{MLP}
    \left(
    \mathbf{x}_{v}^{(k-1)}
    \Vert
    \mathbf{x}_{w}^{(k-1)}
    \right),
    \label{eq:gnn_update}
\end{equation}
where $\mathcal{N}(v)$ is the neighborhood of node $v$ and $\Vert$ denotes feature concatenation. Although the graph topology varies with the problem size, the initial node-feature dimension equals $N$, so the embedding is again required to provide a common feature dimension.

\textbf{GLAN:} GLAN converts the LAP into an edge-labeling problem on a bipartite graph~\cite{liu2022glan}. Each candidate assignment $(i,j)$ is represented by an edge whose initial attribute is the corresponding cost value, i.e., $\mathbf{e}_{ij}^{(0)}=c_{ij}$. GLAN alternates node and edge updates to propagate information over the graph. At each layer, an edge representation is updated from its two endpoint nodes and its previous state:
\begin{equation}
    \mathbf{e}_{ij}^{(k)}
    =
    \phi_{e}
    \left(
    \mathbf{h}_{i}^{(k-1)},
    \mathbf{h}_{j}^{(k-1)},
    \mathbf{e}_{ij}^{(k-1)}
    \right),
    \label{eq:glan_edge_update}
\end{equation}
where $\mathbf{h}_{i}^{(k-1)}$ and $\mathbf{h}_{j}^{(k-1)}$ denote the endpoint-node representations and $\phi_e$ denotes the learnable edge-update function. A decoder then predicts whether each edge belongs to the assignment, trained with a balanced edge-classification loss with soft penalties on the 1-1 constraints. Since each initial edge attribute is a single cost entry, the same network operates on graphs built from cost matrices of any size, and no embedding is required.

The three models expose different representations, but each returns a continuous score matrix $\widetilde{\mathbf{Y}}$, of size $N\times N$ for CNN and GNN and $m\times n$ for GLAN. The CNN discretizes the scores row-wise, consistent with its row-wise classifiers, as
\begin{equation}
    \widehat{y}_{ij}^R = 
    \begin{cases}
        1, & \text{if } \widetilde{y}_{ij} = \max_{k} \widetilde{y}_{ik}, \\
        0, & \text{otherwise},
    \end{cases}
    \label{eq:raw_discretization}
\end{equation}
whereas the GNN takes the maximum over each column instead, and GLAN thresholds each edge score independently. Note that none of these rules enforces the 1-1 constraints. A column may be selected by several rows under the CNN rule, a row may be selected by several columns under the GNN rule, and, under thresholding, a row or column may receive no selection at all. The raw prediction $\widehat{\mathbf{Y}}^R$ is therefore not a feasible mapping in general, and the greedy step is applied to obtain one, as described below.

\subsection{Post-processing: Greedy collision-avoidance rules} \label{sec:post}
\begin{table*}[t]
\centering
\caption{Assignment accuracy of CNN and GNN with (w/) and without (w/o) the embedding across input sizes.}
\label{tab:cnn_gnn_result}
\begin{subtable}[t]{0.49\textwidth}
    \centering
    \caption{CNN}
    \label{tab:cnn_result}
    \renewcommand{\arraystretch}{1.2}
    \begin{tabular}{llcccc}
    \toprule
    Metric & Method & $2\times2$ & $4\times4$ & $16\times16$ & $64\times64$ \\
    \midrule
    \multirow{2}{*}{$A_o^R\uparrow$} & w/o embedding & 99.7 & 74.1 & 51.4 & \textbf{56.3} \\
    & w/ embedding & \textbf{100.0} & \textbf{93.7} & \textbf{75.8} & 41.3 \\
    \hline
    \multirow{2}{*}{$A_o^P\uparrow$} & w/o embedding & 90.0 & 61.3 & 33.4 & \textbf{32.4} \\
    & w/ embedding & \textbf{90.4} & \textbf{77.3} & \textbf{50.9} & 26.1 \\
    \hline
    \multirow{2}{*}{$A_f^R\uparrow$} & w/o embedding & 99.6 & 61.4 & 23.9 & \textbf{12.4} \\
    & w/ embedding & \textbf{100.0} & \textbf{90.2} & \textbf{41.1} & 9.0 \\
    \hline
    \multirow{2}{*}{$A_f^P\uparrow$} & w/o embedding & 87.6 & 47.2 & 10.5 & \textbf{4.8} \\
    & w/ embedding & \textbf{88.0} & \textbf{67.6} & \textbf{21.6} & 4.3 \\
    \hline
    \multirow{2}{*}{$V^R\uparrow$} & w/o embedding & 75.2 & 51.6 & 22.4 & \textbf{11.5} \\
    & w/ embedding & \textbf{75.6} & \textbf{56.6} & \textbf{25.0} & \textbf{11.5} \\
    \bottomrule
    \end{tabular}
\end{subtable}
\hfill
\begin{subtable}[t]{0.49\textwidth}
    \centering
    \caption{GNN}
    \label{tab:gnn_result}
    \renewcommand{\arraystretch}{1.2}
    \begin{tabular}{llcccc}
    \toprule
    Metric & Method & $2\times2$ & $4\times4$ & $16\times16$ & $64\times64$ \\
    \midrule
    \multirow{2}{*}{$A_o^R\uparrow$} & w/o embedding & 99.0 & 89.9 & 63.9 & \textbf{3.5} \\
    & w/ embedding & \textbf{100.0} & \textbf{94.6} & \textbf{74.3} & \textbf{3.5} \\
    \hline
    \multirow{2}{*}{$A_o^P\uparrow$} & w/o embedding & 97.7 & 78.9 & 48.7 & \textbf{2.3} \\
    & w/ embedding & \textbf{99.9} & \textbf{93.7} & \textbf{72.0} & \textbf{2.3} \\
    \hline
    \multirow{2}{*}{$A_f^R\uparrow$} & w/o embedding & 98.7 & 82.5 & 29.1 & \textbf{2.3} \\
    & w/ embedding & \textbf{100.0} & \textbf{91.4} & \textbf{38.4} & 2.1 \\
    \hline
    \multirow{2}{*}{$A_f^P\uparrow$} & w/o embedding & 97.1 & 69.3 & 16.2 & \textbf{0.3} \\
    & w/ embedding & \textbf{99.9} & \textbf{91.9} & \textbf{40.2} & 0.2 \\
    \hline
    \multirow{2}{*}{$V^R\uparrow$} & w/o embedding & 75.2 & 57.6 & \textbf{25.0} & \textbf{2.9} \\
    & w/ embedding & \textbf{75.6} & \textbf{58.2} & 24.9 & 2.1 \\
    \bottomrule
    \end{tabular}
\end{subtable}
\end{table*}

Each model resolves the conflicts in $\widehat{\mathbf{Y}}^R$ with its own greedy rule, yielding a feasible mapping $\theta\in\Theta$ represented by $\widehat{\mathbf{Y}}^P$, which has the same size as $\widetilde{\mathbf{Y}}$.
\begin{itemize}
    \item \textbf{CNN:} For a column selected by several rows, the pair with the lowest original cost $c_{ij}$ is retained, and the conflicting rows are greedily reassigned to unused columns in order of increasing cost~\cite{lee2018deep_lsap}.
    \item \textbf{GNN:} For a row selected by several columns, the pair with the highest score $\widetilde{y}_{ij}$ is retained, and the conflicting columns are greedily reassigned to unused rows in order of decreasing score~\cite{aironi2024graph}.
    \item \textbf{GLAN:} High-confidence non-conflicting edges are locked first, and the unmatched rows and columns are then paired by increasing original cost~\cite{liu2022glan}.
\end{itemize}

\section{Experimental Results}
\label{sec:Results}

This section evaluates the three DL-based LAP solvers, first on synthetic LAP instances and then within a GLMB tracker.

\subsection{Synthetic linear assignment problems}
\begin{table*}
    \centering
    \renewcommand{\arraystretch}{1.2}
    \caption{Assignment accuracy and inference time of the LAP solvers on randomized cost-matrix shapes.}
    \label{tab:solver_comparison}
    \begin{tabular}{lllllll}
        \toprule
        Solver & $A_o^R\uparrow$ & $A_o^P\uparrow$ & $A_f^R\uparrow$ & $A_f^P\uparrow$ & $V^R\uparrow$ & Inference time ($\mu$s)$\downarrow$ \\
        \midrule
        CNN & 41.3 & 26.1 & \textbf{9.0} & 4.3 & \textbf{11.5} & 968.7 \\
        GNN & 3.5 & 2.3 & 2.1 & 0.2 & 2.1 & \textbf{105.9} \\
        GLAN & \textbf{74.0} & \textbf{71.8} & 5.1 & \textbf{6.5} & 10.4 & 139.8 \\
        \hline
        Hungarian Python & 100.0 & 100.0 & 100.0 & 100.0 & 100.0 & 42549.4 \\
        Hungarian C++ & 100.0 & 100.0 & 100.0 & 100.0 & 100.0 & \textbf{32.9} \\
        \bottomrule
    \end{tabular}
\end{table*}

\subsubsection{Experimental setup}
\label{sec:experimental_setup}

We evaluate the three solvers on a synthetic dataset of 500,000 randomly generated cost matrices, divided into training, validation, and test subsets with a ratio of 0.7:0.2:0.1. The data-generation and input-processing procedures are shared by CNN, GNN, and GLAN, for which we keep the default settings. All experiments run on an AMD Ryzen 9970X CPU with 64~GB of RAM and an NVIDIA RTX PRO 6000 Blackwell GPU with 96~GB of VRAM, using Python and PyTorch. The reported inference time of each neural solver covers its full pipeline, namely the embedding (for CNN and GNN), the forward pass, the discretization, and the post-processing.

Accuracy is measured over $T$ testing samples, where sample $t$ has a cost matrix of dimensions $m_t\times n_t$. Let $\mathbf{Y}^{t}$ and $\widehat{\mathbf{Y}}^{R,t}$ denote the ground-truth and raw predicted assignments of sample $t$, and let $\Pi_t=\Pi_{m_t,n_t}$ denote the projection in Eq.~\eqref{eq:crop}. The $i$-th rows of $\Pi_t(\mathbf{Y}^{t})$ and $\Pi_t(\widehat{\mathbf{Y}}^{R,t})$, denoted by $\mathbf{y}_{i}^{t}$ and $\widehat{\mathbf{y}}_{i}^{R,t}$, indicate the measurement selected for object $i$. The per-object accuracy and the full-assignment accuracy of the raw prediction are
\begin{equation}
    \begin{aligned}
    A_o^R &= \frac{1}{T}\sum_{t=1}^{T}\frac{1}{m_t}\sum_{i=1}^{m_t}\mathbf{1}\!\left[\widehat{\mathbf{y}}_{i}^{R,t}=\mathbf{y}_{i}^{t}\right],\\[2pt]
    A_f^R &= \frac{1}{T}\sum_{t=1}^{T}\mathbf{1}\!\left[\Pi_t\big(\widehat{\mathbf{Y}}^{R,t}\big)=\Pi_t\big(\mathbf{Y}^{t}\big)\right],
    \end{aligned}
    \label{eq:assignment_accuracy}
\end{equation}
where $\mathbf{1}[\cdot]$ is the indicator function. Their counterparts after post-processing, $A_o^P$ and $A_f^P$, follow the same expressions with $\widehat{\mathbf{Y}}^{P,t}$ in place of $\widehat{\mathbf{Y}}^{R,t}$. In both, a cropped prediction with fewer than $\min(m_t,n_t)$ original pairs counts as incorrect. The valid-assignment ratio measures how often the raw prediction is already feasible. Unlike the accuracies, it is evaluated before cropping, so for CNN and GNN it covers the full $N\times N$ matrix including the padded rows and columns. It is defined as
\begin{equation}
    V^R=
    \frac{1}{T}\sum_{t=1}^{T}\mathbf{1}\!\left[
    \widehat{\mathbf{Y}}^{R,t}\in\Theta
    \right].
    \label{eq:valid_ratio}
\end{equation}
A raw prediction can therefore agree with the ground-truth on the original block while its padded part violates the 1-1 constraints, so $V^R$ may fall below $A_f^R$. The post-processing of Section~\ref{sec:post} returns a feasible mapping for every instance, so the corresponding ratio $V^P$ equals $100\%$ for all solvers and is not reported in the tables.

Two scenarios are considered. The first compares CNN and GNN with the embedding (proposed) and without it (baseline) at input sizes $2$, $4$, $16$, and $64$. Without the embedding, a separate model is trained at each fixed input size. With the embedding, a single model is trained on rectangular instances whose row and column dimensions vary independently. Both are evaluated on rectangular test instances at each input size. This setting examines whether a model trained on variable-size instances retains assignment quality while supporting non-uniform input dimensions. The second scenario compares the three solvers under a common LAP setting, with CNN and GNN using the embedding at $N=64$ and GLAN processing each matrix at its original size. The largest target size trained in this study is $N=64$, which covers the cost matrices of our real-world tracking experiment in Section~\ref{sec:WT}, and a larger $N$ requires retraining but no change to the embedding.

\subsubsection{Results}

Tab.~\ref{tab:cnn_gnn_result} reports CNN and GNN without and with the embedding. For both models, the embedding generally improves assignment accuracy at small and medium input sizes, with the clearest gains at $4\times4$ and $16\times16$ across $A_o^R$, $A_o^P$, $A_f^R$, and $A_f^P$. However, at $64\times64$, the CNN with the embedding falls below the one without it for most metrics. Both GNN settings also degrade sharply at this size, with $A_o^R$ near $3.5$ and $A_f^P$ approaching zero. These results indicate that the current CNN and GNN architectures and training settings do not scale reliably to larger LAP instances.

Tab.~\ref{tab:solver_comparison} summarizes the assignment accuracy and inference time. GLAN achieves the highest per-object accuracy among the three networks, with $A_o^R=74.0$ and $A_o^P=71.8$, whereas CNN obtains the highest $A_f^R$ and GNN again performs poorly at $N\!=\!64$. The embedding lets a single CNN or GNN model process every size up to $N$, but the edge-based input of GLAN, which needs no embedding, retains more accuracy at this size. In inference time, the optimized Hungarian C++ implementation is the fastest at $32.9~\mu$s, GNN is the fastest network at $105.9~\mu$s, followed by GLAN at $139.8~\mu$s and CNN at $968.7~\mu$s. All three networks are therefore faster than the pure Hungarian Python implementation but slower than the optimized C++ solver at this size. These times include the post-processing, and the current greedy rules of all three networks become increasingly expensive as the matrix size grows, so further acceleration, in particular of the post-processing, is required to reach the speed of the C++ solver. 

\subsection{A real-world tracking problem} \label{sec:WT}

\subsubsection{Experimental setup}

The three solvers are integrated into a GLMB tracker~\cite{Vo2017} and evaluated on the WILDTRACK dataset~\cite{Chavdarova2018}, which contains 7 synchronized cameras observing the same crowded outdoor area. The 3DROM detector~\cite{Qiu2022} fuses the per-camera detections into ground-plane $(x,y)$ coordinates. We use the standard linear Gaussian constant-velocity model in~\cite{Vo2014}, with a kinematic state of planar position and velocity and measurements that are noisy observations of the position. The process and measurement noise standard deviations are set to $0.15$~m/s$^2$ and $0.1$~m, respectively. Since 3DROM is trained on the first 90\% of the sequence, every tracker in Tab.~\ref{tab:WT_pef} is evaluated on the final 40 frames, which form the remaining 10\%. The GLMB filter produces the likelihood scores that define the measurement-to-object assignment cost, and all four GLMB trackers share the same detections and filter configuration, so they differ only in the solver applied to these costs. The costs vary in size across frames. For CNN and GNN, they are embedded with a target size of $N=64$, which covers the largest numbers of objects and measurements observed per frame, whereas GLAN processes them at their original size. Tracking performance is evaluated via \texttt{py-motmetrics} toolbox~\cite{pymotmetrics} with a $1$~meter threshold on the ground-plane for all trackers, reporting the numbers of false positives (FP), false negatives (FN), and identity switches (IDs), the identity F1 score (IDF1), and the multiple-object tracking accuracy (MOTA).

\subsubsection{Results}
\begin{table}
\centering
\caption{Tracking performance with the 3DROM~\cite{Qiu2022} detector on the WILDTRACK dataset.}
\label{tab:WT_pef}
\renewcommand{\arraystretch}{1.2}
\begin{tabular}{llllll} 
\toprule
Tracker & FP$\downarrow$ & FN$\downarrow$ & IDs$\downarrow$ & IDF1$\uparrow$ & MOTA$\uparrow$\\ 
\hline
GLMB CNN & \textbf{23} & 751 & 92 & 14.6 & 9.0  \\ 
GLMB GNN & 45 & 689 & 152 & 13.7 & 6.9   \\ 
GLMB GLAN & 48 & 82 & 23 & 88.4 & 83.9  \\ 
GLMB Hungarian~\cite{Vo2017} & 30 & 65 & \textbf{4} & 94.4 & 89.6   \\ 
\hline
ReST~\cite{Cheng2023} & 75 & \textbf{54} & 29 & 85.9 & 83.4 \\
EarlyBird~\cite{Teepe2024a} & 42 & 63 & 6 & 92.8 & 88.3 \\
TrackTacular~\cite{Teepe2024b} & 31 & 57 & 7 & \textbf{94.5} & \textbf{90.0} \\
\bottomrule
\end{tabular}
\end{table}

Tab.~\ref{tab:WT_pef} reports the tracking performance of the four solvers within the same GLMB tracker, together with three state-of-the-art DL-based trackers re-run from their official implementations. ReST~\cite{Cheng2023}, EarlyBird~\cite{Teepe2024a}, and TrackTacular~\cite{Teepe2024b} consume the raw per-camera frames and fuse them within an E2E model to output trajectories directly, whereas the LAP solvers studied here operate inside a modular pipeline and replace only the assignment step. Among the three networks, GLAN transfers best to the tracking task, reaching an IDF1 of $88.4$ and a MOTA of $83.9$, close to the Hungarian reference at $94.4$ and $89.6$. CNN and GNN produce many missed detections and identity switches, and their IDF1 and MOTA remain low. Overall, these results demonstrate the feasibility and potential of DL-based LAP solvers within a tracker, with GLAN already approaching the Hungarian reference, although a gap remains.

\section{Conclusions}\label{sec:Conclusion}

This paper studied DL-based LAP solvers as a replacement for the Hungarian algorithm in MOT. A cost-matrix embedding was introduced so that a single model handles cost matrices of any size up to its target dimension, and CNN, GNN, and GLAN were evaluated on synthetic LAP instances and within a GLMB tracker on WILDTRACK. The embedding improved the accuracy of CNN and GNN at small and medium sizes, GLAN attained the best accuracy among the three, and, within the tracker, GLAN approached the Hungarian reference while the neural solvers ran faster than the pure Python implementation. These results demonstrate the feasibility and potential of DL-based LAP solvers for tracking. Future work will explore other neural models to improve the assignment accuracy, accelerate the post-processing to reach the speed of optimized exact solvers, and exploit the differentiable score matrix to train end-to-end trackers.

\bibliographystyle{IEEEtran}
\bibliography{reference.bib}	
\end{document}